\documentclass[12pt,preprint]{aastex}

\def\amax{$a_{\rm max}$}
\def\amin{$a_{\rm min}$}
\def\lam{$\lambda$}
\def\mtex{$\langle\tau_{\rm ext}\rangle$}
\def\mtsca{$\langle\tau_{\rm sca}\rangle$}
\def\ngc{NGC\,7023}
\def\tex{$\tau_{\rm ext}$}

\def\tpm{two-phase models}
\def\tsca{$\tau_{\rm sca}$}

\def\tz{$\tau_0$}

\begin{document}
\title{Can Reflection from Grains Diagnose the Albedo?}
\author{John S. Mathis\footnote{Univ. of Wisconsin-Madison,
475 N. Charter St.,
Madison WI 53706; mathis@astro.wisc.edu}}
\author{Barbara A. Whitney\footnote{Space Science Institute,
3100 Marine Street, Suite A353,
Boulder, CO 80303; bwhitney@colorado.edu}}
\author{Kenneth Wood\footnote{School of Physics and Astronomy,
University of St. Andrews, North Haugh,
St. Andrews, Fife KY16 9AD, Scotland,
UK; kw25@st-andrews.ac.uk}}

\begin{abstract}
By radiation transfer models we show that the optical properties of
grains are poorly constrained by observations of reflection nebulae.
The interstellar medium is known to be hierarchically clumped from a
variety of observations (molecules, H~I, far-infrared). We have
performed radiative transfer through four-tiered hierarchically
clumped dust in a sphere surrounding a central star. Our models have
realistic power spectra of the projected density distributions (index
$\sim-3$). The input parameters are the albedo ($a$) and phase
parameter ($g$) of the dust, the radial optical depth of the sphere
averaged over all directions (\tz), and the detailed random
distribution of the dust clumps within the sphere. The outputs are the
stellar extinction, optical depth, and flux of scattered light as seen
from various viewing angles. Observations of a reflection nebula
provide the extinction and scattered flux as viewed from one
particular direction.

Hierarchical geometry has a large effect on the flux of scattered
light emerging from a nebula for a particular extinction of the
exciting star. There is a very large spread in both scattered fluxes
and stellar extinctions for any distribution of dust. Consequently, an
observed (\tex,\,\tsca) can be fitted by a wide range of albedos.

There are lower limits on $a$ set by the scattered flux. As an
example, in the best observed reflection nebula, \ngc, $a$(1300\,\AA)
must be larger than $\sim$\,0.5 if the scattered flux from Witt et al
(1993) and a reasonable value for the optical depth within the nebula
are adopted. However, the same observations can be fitted with $a$ =
0.8 and $0.6\le g\le 0.85$, the entire range we considered.

With hierarchical geometry it is not completely safe to determine even
relative optical constants from multiwavelength observations of the
same reflection nebula. The problem is that the geometry effectively
changes with wavelength as the opacity of the clumps varies. Limits on
the implications of observing the same object in various wavelengths
are discussed briefly.

Henry (2002) uses a recipe to determine the scattered flux from a star
with a given extinction. It is claimed to be independent of the
geometry. It provides considerably more scattering for given dust
optical properties than our models, probably leading to an
underestimate of the grain albedos from the UV Diffuse Galactic Light.

\end{abstract}

\section{Introduction}

Interstellar dust is poorly understood in spite of its importance
within galaxies. The mean extinction law (i.e. scattering plus
absorption) for the diffuse Galactic interstellar medium (ISM) is
reasonably well known, but the individual absorption and scattering
properties are still controversial. Many other basic properties
(chemical composition, particle size distribution, particle shape
distribution, alignment of the grains in the magnetic field, and
others) are also poorly known.

The properties of dust for each scattering event are characterized by
two parameters: $a(\lambda)$, the albedo or fraction of the extinction
that is scattering, and $g(\lambda)$, the mean cosine of the
scattering angle, averaged over the phase function for single
scattering. Geometrical uncertainties prevent the determination of
higher moments of the scattering pattern beyond these two. The $a$ and
$g$ are estimated from either reflection nebulae excited by a central
star, radiation from a dusty globule reflecting Galactic light, or the
``diffuse galactic light" (DGL), the general stellar radiation
reflected by the dust distributed throughout the ISM.

Almost always, nebulae have been interpreted with smooth (non-clumpy)
models, sometimes with a density gradient. Radiation transfer differs
in smooth and clumpy media because photons can escape between the
clumps. Uniformity was a reasonable assumption, for three reasons.
(a) It is the simplest, and Occam's Razor has great appeal. (b)
Hierarchical clumping had not been convincingly demonstrated by
several means. (c) The computational resources required for radiative
transport through hierarchical clumps have only recently become widely
available.

Any estimate of the albedo is affected by the hierarchical clumping of
the ISM. These clumps occur on scales from AU to tens or hundreds of
parsecs, as observed in various molecules (CO, formaldehyde, and OH),
ionic species, H~I, and 100$\mu$m (Crovisier \& Dickey 1983; Green
1993; Moore \& Marscher 1995; Vogelaar \& Wakker 1994; Elmegreen \&
Falgarone 1996; Abergel et al. 1996; Elmegreen 1997; Heiles 1997;
Heithausen et al. 1998; Falgarone et al. 1998; Chappell \& Scalo 2001;
Welty \& Fitzpatrick 2001; Faison \& Goss 2001; Andrews, Meyer, \&
Lauroesch 2001). Other galaxies (Stanimirovi\'{c} et al. 1999 for the
SMC; Westpfahl et al. 1999 for the M81 group) show the same phenomena
in H~I. The best observed reflection nebulae are clumpy
(e.g. Sellgren, Werner, \& Dinerstein 1992; Martini, Sellgren, \&
DePoy 1999; Knauth et al. 2001). We have examined the images in the
SIMBAD database of the reflection nebulae with which we are acquainted
(about a dozen); all appear to be clumpy. The observed thermal
pressure in the ISM as judged from C~I (Jenkins \& Tripp 2001) varies
by over an order of magnitude within a sample of stars against which
the column densities of C~I lines can be determined. The hierarchical
structure of the ISM is a natural consequence of turbulence, which is
scale-free. Turbulent models of the ISM generate hierarchical density
structure (e.g., Norman \& Ferrera 1996).

Boiss\'{e} (1990) determined the transport of radiation through a
two-phased clumpy medium with isotropic scattering. Witt \& Gordon
(1996) extended the calculations to anisotropic scattering in a
centrally illuminated sphere. They investigated the effects of
changing various parameters such as the contrast between the density
contrast between the phases or the sizes of the clumps relative to the
radius of the sphere. Many of the qualitative effects we find from
varying parameters for hierarchical models are discussed by them as
well. As the volume fraction of the dense phase was increased, their
models varied from widely separated dense clumps up to continuous
structures with small holes. Since all cells have either of two
densities, we will refer to this type of clumping as ``two-phase
models''. By contrast, our hierarchical models have an almost
continuous distribution of densities.

Witt \& Gordon (2000) have investigated radiative transfer in galaxies
by means of a specific two-phased ISM, one being 100 times as dense as
the other and occupying 15\% of the volume. In order to compare \tpm\
with hierarchical, we adopt this recipe, with 16 cells along the
radius of the sphere (Witt \& Gordon used 15). All of our \tpm{}
assumed ($a,\,g)$ = (0.6,\,0.6), which are values typical for optical
wavelengths suggested by observations (see plots in Witt \& Gordon
2000).

The fact that the ISM is {\it hierarchically} clumped is important for
the propagation of scattered light, since such a structure has
relatively open spaces through which the radiation can move rather
freely. The main thrust of this paper is not to try to put limits on
the grain properties of real reflection nebulae, but to show the {\it
variation} among hierarchical models as they are viewed from various
angles. We will not be very concerned with the averaged properties of
the models, since each real object is viewed from only one direction.

\section{Hierarchically Clumped Models}

The density structure of the ISM is often described as ``fractal",
meaning self-similar on all size scales. We will consider models that
are hierarchically clumped instead, meaning that they are self-similar
over a limited range (about a factor of ten) in sizes. We use a
procedure similar to that in Elmegreen (1997): (a) Consider a
``supercube", a portion of which will represent a spherical reflection
nebula. The supercube of size L on a side consists of 64 cubical cells
stacked along each dimension. For each cell we determine the local
density of dust, as explained below. (b) Place $N$ points randomly
within the supercube. We used $N$ = 32. (c) Randomly cast another $N$
points, all within a distance $L/(2\Delta)$ in each Cartesian axis
from each of the points cast in the preceding round. Here the distance
$\Delta$ is related to the ``fractal dimension'', $D$, by
$D\equiv\log(N)/\log(\Delta)$. Allow any of the points that fall
outside of the supercube to remain there. (d) Repeat procedure (c),
above, twice more, so that the total number of points cast is
$N^4$. (e) Shift the points outside of the supercube to within it by
translating each Cartesian coordinate outside of the supercube by $L$
until it lies within. This procedure reflects the points so that if
they originally fall off the left side of the supercube they reappear
the same distance from the right within the supercube, and
correspondingly for up/down and front/back. The density within the
supercube is then proportional to the number of points within each
cell. (f) Inscribe a sphere within the supercube and place a point
source of radiation at its center. The constant of proportionality
between the number of points in each cell and the optical depth within
the cell is chosen to make the radial optical depth, averaged over all
directions, be \tz.

The main parameter of this procedure is $D$, which describes the
degree of clumping. We considered $D$ = 2.3 and 2.6, in the range
observed in actual clouds. Another observable parameter is $\beta$,
the exponent of the power of the projection of the density within the
supercube onto the plane of the sky\footnote{ The structure function
$F(\delta{\bf r})$ is defined by $F(\delta{\bf r})\equiv{\rm
(area)}^{-1}\,\int\int\,I({\bf r}) I({\bf r}+\delta{\bf r})\,dx\,dy$,
and $P(k)$ is its Fourier transform, with $k=|\delta{\bf
r}|^{-1}$. Here ${\bf r}$ and {\bf$\delta$r} are vectors, and
($x,\,y$) Cartesian coordinates, in the plane of the sky.}, so that
$P(k)\propto k^{-\beta}$, where $k$ is the wavenumber. Observations
(e.g. Stanimirovi\'{c} et al. 1999) show $\beta\sim3$. With our recipe
we find $\beta=2.8$ with $D$ = 2.6 and $\beta\sim2.3$ with $D$ =
2.3. However, none of the conclusions in this paper depend upon which
set of hierarchical models we consider. Many different models can have
the same $\beta$. No one parameter, either $D$ or $\beta$, can
completely describe even the projection of the density distribution of
the ISM onto the sky.

The above procedure with the parameters we adopted results in clouds
that are quite full of holes, with $\sim$15\% of the projected density
distribution being almost blank. The large-scale ISM as shown by H~I
is similar if a low column density of H~I ($<5\times10^{19}$ H atoms
cm$^{-2}$) is taken to be ``blank'' (i.e., a column density much
smaller than the average). Elmegreen (1997) has discussed the
emptiness of the ISM, with the large-scale structure in mind. Images
of actual reflection nebulae (see Sellgren, Werner, \& Dinerstein 1992
for NGC\,2023 and NGC\,7023) show strong filaments and structure, but
there is material at all points within the nebula. For this reason, we
sometimes add a uniform density to all cells in the supercube. This
constant density does not change the value of $\beta$, since the
Fourier transform of a constant is a Dirac $\delta$-function at the
origin, $k=0$. Choosing a constant density provides a minimum
projected density.

The detailed placement of the clumps of dust also influences the
results of our radiative transfer calculations. The clumping is mainly
formed by the first round of random casting of 32 points that are
subsequently spread by three more rounds of casting points in their
vicinities. The locations of the first round of points is uniquely
determined by the integer seed of the random number
generator\footnote{We used ran2(iseed) as described in Press et
al. (1992).}, and the locations of all subsequent points follow from
the initial value of this integer in a complicated but unique way. The
response of the model nebula to the central star is strongly affected
by the precise placement of the dust relative to the star and,
thereby, to the value of the initial seed. We considered suites of
models that differed only in their initial seeds. Our procedure
amounts to assuming that the star does not affect the density
distribution of the dust in its immediate vicinity, so the initial
seed is allowed to control the placement of the dust both near and far
from the star. To test the sensitivity of results to this assumption,
we also considered models with a cavity in the dust distribution
within the inner 10\% of the radius of the sphere.

The radiative transfer was performed by the Monte Carlo code described
by Wood \& Reynolds (1999). It involves considering photon packages
propagated in a random direction from the central star. We assumed $a$
and $g$, along with the scattering phase function of Henyey \&
Greenstein (1941). In most cases we used $5\times10^6$ photons for
each model, after checking for a few cases that the results were the
same as from a model using $2\times10^7$ photons. Unlike the situation
for placing the dust, the initial seed for the radiative transfer
makes no difference, as expected from such a large number of stellar
photons propagated in random directions.

We calculated the fluxes of scattered radiation and also of starlight,
both of which depend on the angle from which the sphere is viewed
because of the strong randomly placed clumping. Both fluxes are
expressed relative to the stellar flux that would be observed from the
unobscured star. We express the fluxes as extinction and scattering
optical depths, \tex{} $\equiv-{\rm ln}(F_*/F_{*0})$ and \tsca{}
$\equiv-{\rm ln}(F_{\rm sca}/F_{*0})$, where $ F_{*0}$ is the flux the
star would have if there were no reflection nebula. Each flux or
optical depth depends on the direction from which the sphere is
viewed. We considered 18 evenly spaced values of cos($\theta$) and 36
values of $\phi$, where $\theta$ is the polar angle of the sphere and
$\phi$ the azimuthal. $F_{\rm sca}$ and $F_*$ can be determined
directly (with difficulty!) in real nebulae. $F_{*0}$ can be
determined if there is enough wavelength coverage to determine the
extinction from extrapolation to large wavelengths, from colors if the
reddening law is assumed, or if the total FIR flux from the nebula is
measured. Otherwise, the ratio of the stellar to the nebular flux
gives the difference of \tex{} and \tsca{}. We now ask, how well can
we determine ($a,\,g$) from an observation of (\tex,
\tsca)?

\section{Results} 

As regards radiative transfer, hierarchical clumping introduces major
differences from a uniform distribution or, to a lesser extent, from
\tpm. The main properties of hierarchical models are:

(a) When viewed from various directions, hierarchically clumped
nebulae exhibit \tex{} and \tsca{} spanning a wide range. A real
reflection nebula represents viewing a collection of dust from one
particular direction. We can calculate mean properties of models as
averages over all viewing directions, but mean properties are relevant
only if a large collection of nebulae (such as in another galaxy) are
observed. Angle-averaged properties of any model have very limited
applicability for interpreting a given reflection nebula. The \tpm{}
models of Witt \& Gordon (1996) also show a wide range of \tex{} at
a given \tz.

(b) As expected, the global parameters of the models, such as
\tz{} and $D$, have a significant influence on the radiation transfer.
We considered both purely hierarchical models and some with 33\% of
the mass in a constant density sphere upon which the clumps are
superimposed. This uniform background density can have an important
effect on the radiation transfer in optically thick nebulae, when the
diffusion depends more upon the transparent regions than on the
opaque. This rule is familiar in connection with the use of the
Rosseland mean opacity in the optically thick limit of radiation
diffusion in stars. Whether such a uniform layer exists in reflection
nebulae is not clear. The regions between clumps on large (molecular
cloud) scales have a vastly lower density than the mean (e.g.,
Elmegreen 1997).

(c) Models with the same global parameters (\tz, $D$, $a$, $g$), but
differing in the placement of the dense clumps relative to the central
star, show strong variations in even the averaged fluxes. There are
significant differences between hierarchical models in which the star
happens to lie within a dense clump of dust, in contrast to those in
which it falls within a void (or if there is a central hole).  If the
star happens to lie within a void, there are lightly reddened paths
that reach from the star to the edge of the sphere. If the star is
almost unattenuated as seen from a particular direction, the nebula
would be interpreted as having the star in front of the dust, except
that there can be a significant amount of scattered light that would
not be there if the star were truly foreground. In these cases, the
albedo derived from uniform models would be $>1$.

The power law index, $\beta$, of the projected density distribution
of our \tpm{} is $\sim$\,0, reflecting the fact that their projected
density is non-hierarchical. The \tpm{} always provide a smaller
range in \tsca($\theta, \phi$) (about $\pm$0.12) than the
hierarchical, but their spread in \tex{} is comparable to
hierarchical models at low \tz{} ($\lesssim2$). The spread at large
\tz{} decreases because the interclump medium becomes opaque,so that
in the ultraviolet (UV) \tpm{} make a prediction of $a$ that is rather
independent of viewing angle.

\begin{figure}
\vspace{-1in}
\plotone{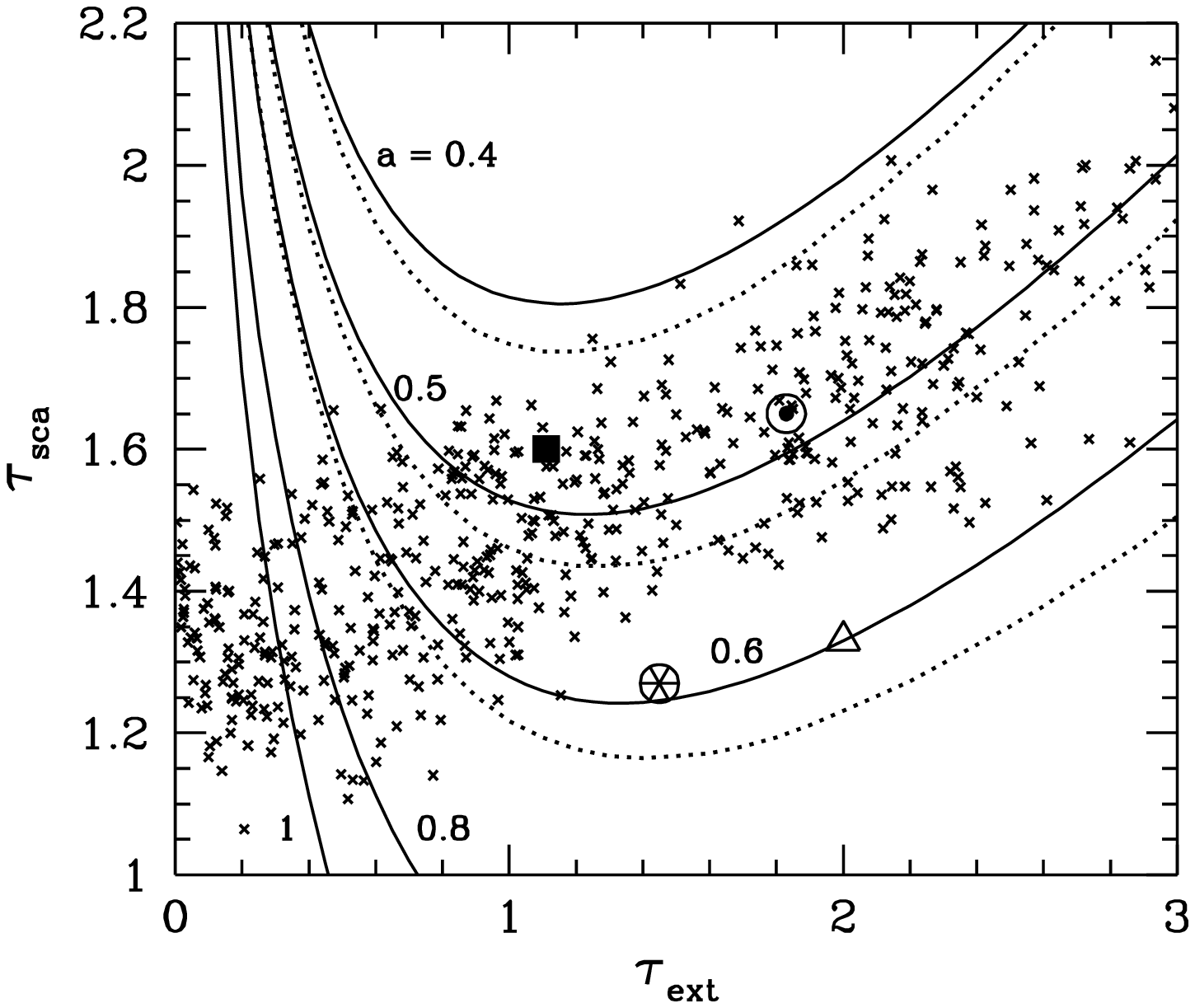}
\vspace{-1.8in}
\caption{The optical depth of scattering in a spherical
cloud illuminated by a central star, plotted against the effective
optical depth for the stellar radiation. Here \tsca{}
$\equiv-\ln[F_{\rm sca}/F_{*0}]$, $F_{\rm sca}$ is the scattered flux
from the nebula, and $F_{*0}$ is the stellar flux if there were no
reflection nebula. Similarly, \tex{}$\equiv-\ln[F_*/F_{*0}]$. The
small crosses represent the fluxes from a hierarchical model with
albedo, $a$, =0.6, phase parameter, $g$, =0.6, average radial optical
depth for the sphere, \tz, = 2, and clumping with a ``fractal
dimension'', $D$, = 2.6. Each small cross represents a direction of
viewing the sphere. The solid square represents (\mtex, \mtsca), the
fluxes (not optical depths) of the model, averaged over viewing angles
(see eqn. 1). We see a very wide range of points for the hierarchical
model as viewed for various directions. The open triangle is the point
for a uniform model. The circled asterisk is the observed point for
\ngc, a well-observed reflection nebula, at 1300
\AA\ (Witt et al. 1993) with 40\% of the observed extinction towards
the central star to be in \ngc{}, and the circled dot is if 50\% of the
extinction is in \ngc. The solid lines give the relation for a uniform
distribution, with various dust albedos marked. The dotted lines show
the relationship assumed by Henry (2002) in analyzing the Diffuse
Galactic Light. The albedos are those indicated at the nearby solid
lines.
\label{fig1}}
\end{figure}

Figure \ref{fig1} shows \tsca{} plotted against \tex, for the case
\tz{} = 2, $D$ = 2.6, $a$ = 0.6, $g$ = 0.6 for a hierarchical
distribution of clumps with an initial seed for the density
distribution that provides a typical (defined below) placement with
respect to the star. Each point represents a particular viewing
direction. One sees a large spread in the values of
\tex. The values for \tex{} for this model range up to 6.8, and
\tsca{} to 2.4. Thus, this one model of clumpy dust can produce a wide
range of stellar fluxes and scattered light.

We represent the direction-averaged escaping flux (not mean optical
depth) by \mtex, defined by 
\begin{equation}
\exp(-\langle\tau_{\rm ext}\rangle)
=\sum_{\theta,\phi}\,\exp[-\tau_{\rm ext}(\theta,\phi)]/
N_{\theta,\phi} ,
\end{equation}
\noindent where $N_{\theta,\phi}$ is the number of bins we are
using ($= 18\times36$), all of equal solid angle. If the star is
embedded within a clump, the uniform dust component is of reduced
importance because scattering occurs within the clump. Similarly,
\mtsca{} represents the angle-averaged scattered radiation, with \tsca{}
replacing \tex{} in the above relation. The filled square in
Figure \ref{fig1} shows \mtex\ and \mtsca, which are useful for
contrasting the results of models with the same value of \tz{} but
different initial seeds. As the figure shows, the (\mtex,
\mtsca) point is of very limited value in guessing what actual optical
depths a given nebula might show if it were viewed from various
angles. In this model \mtex, which describes the average optical depth
of the stellar flux that escapes from the nebula, is only 0.5\tz, the
averaged optical depth through both the clumps and interclump
material. It is sobering to recall that only a handful of reflection
nebulae have been well observed, while each point in the figure
represents a view of a single model with a particular clumping and
$a$, $g$, and \tz.

The solid lines in Figure \ref{fig1} are the tracks of a spatially
uniform dust distribution for the same $g$ (= 0.6) and various values
of $a$, as labeled. These lines were computed with the surprisingly
accurate analytical approximation (the scattered flux is
underestimated by $\le$3\% for \tex\ $<$ 4) given by Code (1973). Most
reflection nebulae have been interpreted using these uniform models,
in which case \tex{} is the true optical depth of the nebula. For our
hierarchical model, in about 15\% of viewing directions we would
conclude that the star is in front of most of the dust because the
uniform models would give $a>1$. In 40\% of the directions we would
conclude that $a>0.6$, the value assumed in the hierarchical model.

Two UV observations of \ngc{} are shown in the figure.  The central
star, HD\,200775, has foreground extinction from ordinary ISM in
addition to the extinction provided by the nebula (see Witt et
al. 1993 for discussion and other references to observations of
\ngc). The circled asterisk marks the nebular parameters derived from
{\it Voyager-2} and Hopkins Ultraviolet Telescope (HUT) data if 40\%
of the total extinction arises within the nebula, and the circled dot
if the nebula provides 50\%. The dashed lines will be discussed in \S4 in
connection with the UV DGL.

\begin{figure}
\plotone{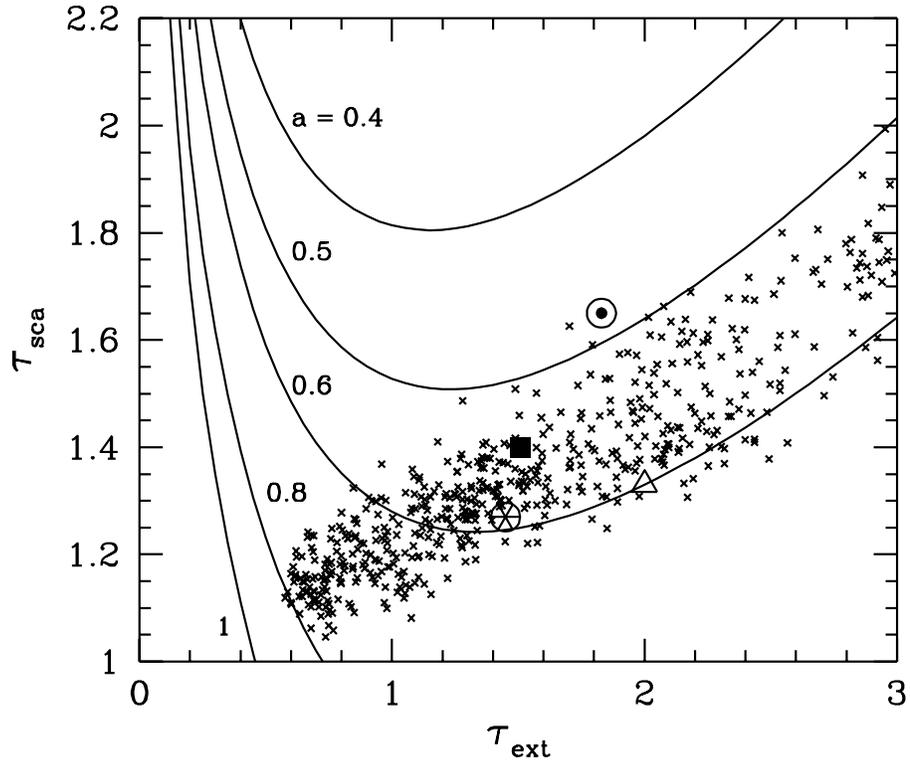}
\caption{The same as in Figure \ref{fig1}, except that 33\% of the
dust density is assumed to a uniform density underlying the
hierarchical clumps. The \tsca{} points are lower than in Figure
\ref{fig1}, showing more scattered light.
\label{fig2}}
\end{figure}

Figure \ref{fig2} is the same model as Figure \ref{fig1}, except 33\%
of the dust is in uniform distribution upon which the hierarchical
clumping is superimposed. The axes of the figures are the same as
Figure \ref{fig1} to emphasize that a uniform density increases the
scattered light, so long as its optical depth along a radius is not
too large ($\lesssim1)$, and imposes a minimum value of \tex{}. The
range of \tsca{} for a given \tex{} is smaller than in Figure
\ref{fig1}, but this is an artifact of the uniformity of the extra
component that is the only difference in the two figures.

\begin{figure}
\plotone{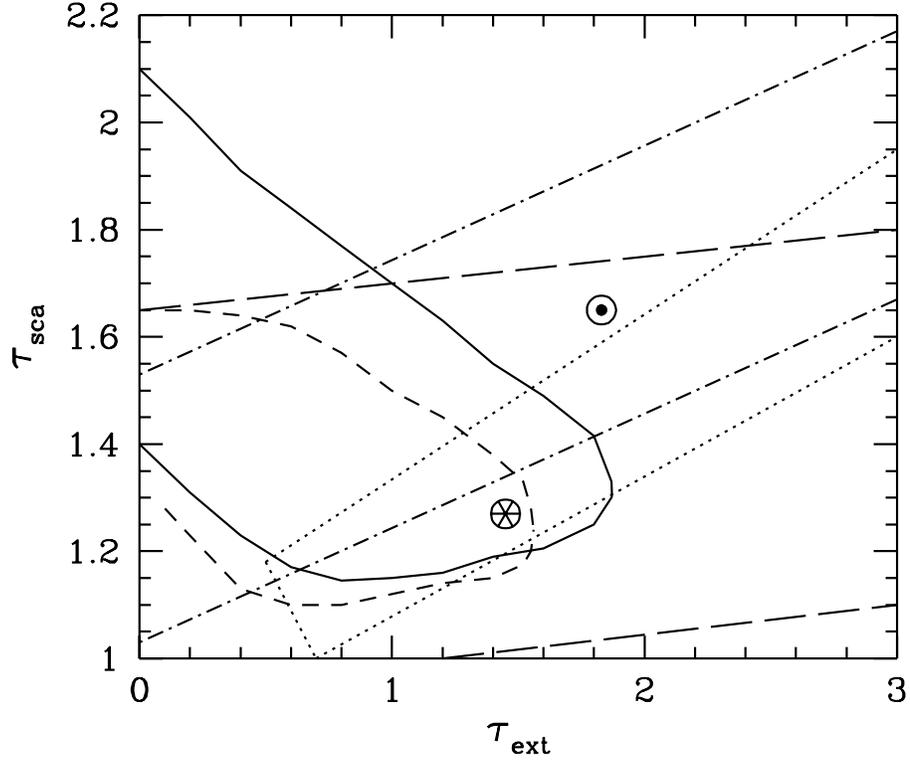}
\caption{
A plot with the same axes as Figures \ref{fig1} and \ref{fig2} showing
the general areas of the (\tex,\,\tsca) plane that are covered by
points from various hierarchical models. ($a,\,g,\,D,\,\tau_0$) values
are: {\it Solid line,} (0.8, 0.6, 2.3, 0.5). {\it Short-dashed line,}
(0.8, 0.6,\,2.3,\,0.5), with 33\% of dust mass in constant
density. {\it Dot-dashed line,} (0.6,\,0.6,\,2.6,\,2). {\it Dotted
line,} (0.6,\,0.6,\,2.6,\,2), with constant density component. {\it
Long dashed line,} (0.8,0.95, 2.3, 4). Marked points: observations for
\ngc\ (see Figures \ref{fig1} and \ref{fig2}). We see that a wide
variety of hierarchical models can cover the observations.
\label{fig3} }
\end{figure}

Wide areas of the (\tex, \tsca) plane are covered by each hierarchical
model, so a given (\tex, \tsca) observation can be covered by a wide
variety of models. By contrast, uniform models have very limited
dependence on $g$, and so contain only one effective parameter,
$a$. The ambiguity of hierarchical models is suggested by
Figure \ref{fig3}. It has the same axes as Figures \ref{fig1} and
\ref{fig2}, but only outlines around the areas covered by the points of
five hierarchical models. The solid line has a high $a$ (= 0.8),
moderate $g$ (0.6), and low \tz{} (0.5). The maximum values of \tex{}
are comparatively small because of the low \tz. The dashed line
encloses the points for the same model with 33\% of the dust mass in a
constant density; the scattered points are lower (there is more
scattering) because there are no almost empty paths through the
nebula. The dot-dashed and dotted lines are the boundaries of the
points in Figures \ref{fig1} and \ref{fig2}, which are for
($a,\,g,\,D,\,\tau_0)$ = (0.6, 0.6, 2.6, 2). The long dashed lines
enclose the very wide boundaries for ($a,\,g,\,D,\,\tau_0)$ = (0.8,
0.85, 2.3, 4). {\it We can have large $g$ or moderate, large $a$ or
moderate, and large \tz{} or small and still fit the
observations}. However, we will see that a minimum $a$ of $\sim$0.5 is
required to produce enough scattered light.

\begin{figure}
\vspace{-1in}
\plotone{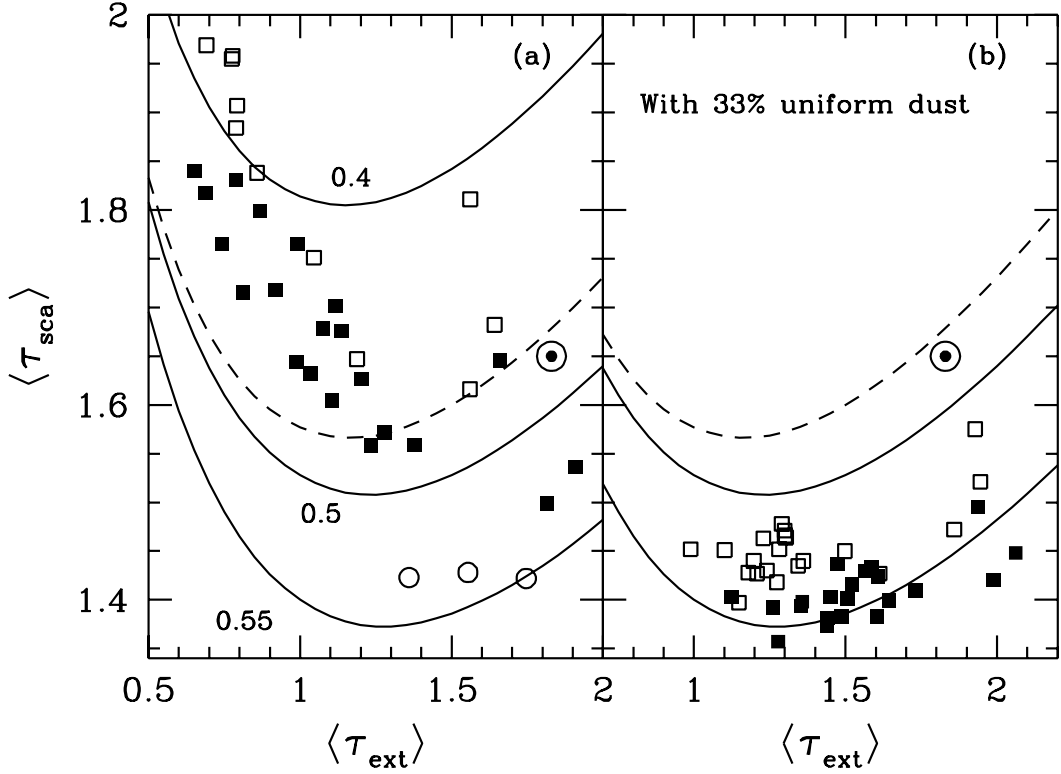}
\vspace{-1in}
\caption{The optical depths for scattering and stellar
radiation for various hierarchical models, averaged over viewing
angles. All have ($a,\,g,\,\tau_0)$ = (0.6, 0.6, 2). Each square
represents a different arrangement of dust clumps. Panel ($a$) shows
models with no uniform component; $b$, models with 33\% of the dust in
a uniform component. Filled squares: $D$ = 2.6; open squares, $D$ =
2.3. Lines: the relationship for spatially uniform dust, with albedo
as marked in panel ($a$), $g$ = 0.6. Open circles in panel ($a$): dust
clumped in either of two densities, with parameters similar to those
in Witt \& Gordon (2000), also with $a$ = 0.6, $g$ = 0.6, and
different random spatial arrangements. The extremes of many different
spatial arrangements are shown, along with a typical case. The dashed
lines show uniform models with $a$ = 0.5 and $g$ = 0.4 instead of 0.6.
We see that the other parameters are more important than $g$.
\label{fig4}}
\end{figure}

Figure \ref{fig4} shows the {\em averaged} (\mtex, \mtsca) values for
all of the 21 different initial seeds that we tried. Figure
\ref{fig4}$a$ shows purely hierarchical models; \ref{fig4}$b$, models
with 33\% of the dust in a constant distribution. The parameters are
the same as in Figures \ref{fig1} and \ref{fig2}: $a$ = $g$ = 0.6,
\tz{} = 2. The \ngc{} points are as before. The open squares are
models with $D$ = 2.3 instead of 2.6. Within each panel, the $D$ = 2.3
models are higher because the radiation, both scattered and stellar,
can escape more easily from the more strongly clumped structure. The
lines show uniform models, with albedos marked in 4$a$. The uniform
model appropriate to the actual albedo assumed in the models ($a$ =
0.6) is just below the bottom of the figure. The dashed line is the
uniform model with $a$ = 0.5 and $g$ = 0.4 instead of 0.6. The
differences are not large in comparison to the effects of the other
parameters. The effects on hierarchical models of changing $g$ are
similar. 

Perhaps the most striking difference between the two panels is the
lower values of \mtsca\ for hierarchical models with 33\% of the dust
in a uniform component. The increase of scattering from the dust
between the clumps causes the decrease in \mtsca{} and greatly reduces
the differences between models with different spatial distributions of
dust. In either panel, the points from various viewing angles of the
two values of $D$ are completely intertwined. Reflection nebulae are
poor diagnostics of $D$ as well as other properties of the ISM.

The three open circles in Figure \ref{fig4} are \tpm{} with three values
of the initial seed. Two of these values show the extrema in (\mtex,
\mtsca) among the 21 initial seeds that we tested. The tightness of 
the mean optical depths of the \tpm{} models shows the importance of
hierarchical geometry, as opposed to simple two-phase clumps. The
\mtsca{} of the \tpm\ models are similar to those of the hierarchical
models with uniform dust. The contrast of both with purely
hierarchical models illustrates the importance of dust-free regions
(in real space, possibly caused by extensions of hot, low-density
material into the nebulae).

Figure \ref{fig4} shows what we meant by saying that Figure \ref{fig1} was
produced by a typical hierarchical model -- one with a typical
\mtex. Those with large \mtex{} have the star embedded within dusty
material, and a relatively low central dust density leads to a low
\mtex.

\begin{figure}

\plotone{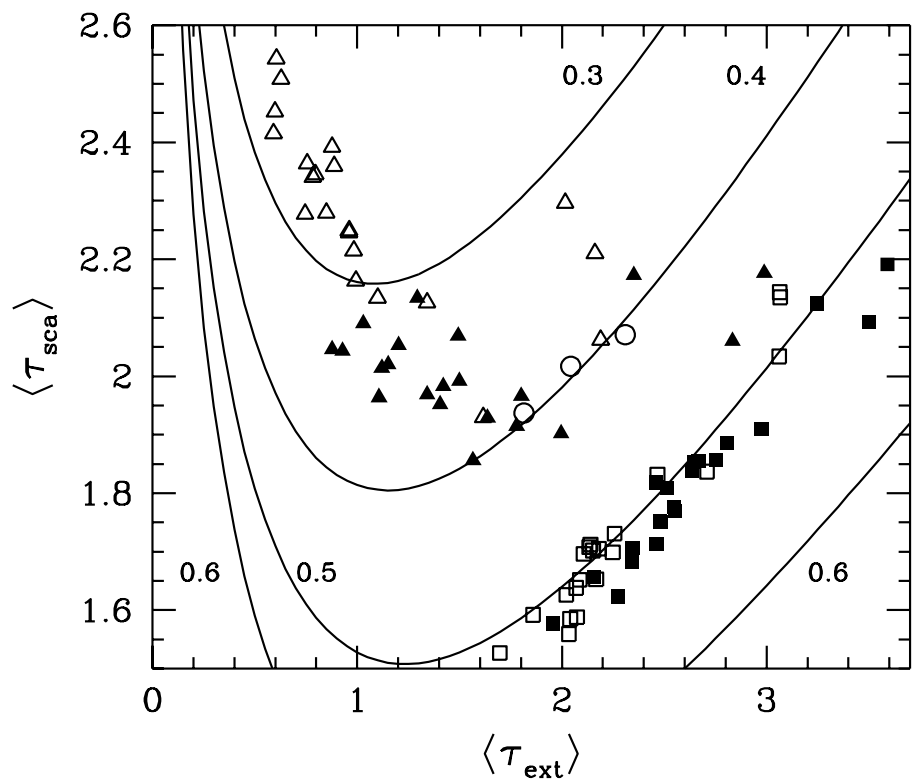}
\caption{The squares are the same as
for Figure \ref{fig4}, 33\% of the density in underlying constant
density, except that the averaged radial optical depth = 4 instead of
2. Open squares: $D$ = 2.3. Filled squares: $D$ = 2.6. Open triangles:
purely hierarchical models (no uniform density), $D$ =2.3. Filled
triangles: purely hierarchical models, $D$ = 2.6. The solid lines are
for spatially uniform dust with $a$ as marked, $g$ = 0.6.  
\label{fig5}}
\end{figure}

Differences among models, and errors arising from interpreting
reflection nebulae with uniform models, increase with optical
depth. Figure \ref{fig5} shows the same as Figure \ref{fig4} ($a$ =
0.6; $g$ = 0.6), except that \tz{}= 4 instead of 2. The open squares
are for various initial dust distributions with $D$ =2.3, with 33\% of
the density uniform; the filled squares, the same with $D$ = 2.6; open
triangles, purely hierarchical density, $D$=2.3; filled triangles,
purely hierarchical with $D$ = 2.6. The circles enclose the entire
range of our \tpm{} models.

Figure \ref{fig5} shows an important result: optically thick
hierarchical models can show very modest values of \mtex, but
interpreting the nebula with uniform models will likely result in a
severe underestimate of the albedo unless there is material between
the clumps to scatter light efficiently. For the purely hierarchical
models with $D$ = 2.3 (open triangles), there is very little scattered
light between the clumps (i.e., large \tsca) and limited stellar
extinction, resulting in albedos predicted from uniform models that
are as low as 0.17. The spread among the \tpm{} (circles) is
relatively small because at large optical depth they become
statistically uniform. Since they are clumpy, the uniform-model
albedos they suggest are significantly lower than the 0.6 used in the
actual \tpm.

The squares in Figure \ref{fig5} have 33\% of the dust mass in a
constant density, so that the minimum optical depth is 4/3. This
addition greatly increases the scattered flux and makes the difference
between $D$ = 2.3 and 2.6 almost unimportant. The albedo that would be
derived from uniform models is increased to $\sim$0.5 over the purely
hierarchical models, still short of the correct value of 0.6.

The spread of the \mtex\ and \mtsca\ in Figures \ref{fig4} and
\ref{fig5} show that the specification of \tz\ and $D$ (along with
grain properties) does not specify the scattered and stellar fluxes
from reflection nebulae in even a statistical sense. By contrast, $D$
constrains the statistical properties of the projected brightness
distribution for both line and continuum emission from the gas. The
large variations in properties of reflection nebulae with viewing
angle come about because the radiation arises from the geometry of the
statistical distribution of the ISM relative to a point source rather
than to itself. Emissions form gas involve an integration over larger
regions of more equal weight.

It is unrealistic to suppose that reflection nebulae are spherical
with a central exciting star.  We have considered only this geometry
because we feel that it provides rather general results for strongly
forward-throwing scattering (say, $g\gtrsim0.5$). In this case, the
dust on the near side of the nebula is likely to provide most of the
scattered light.  We consider our models as representing only this
near-side dust.

\section{Discussion}
\subsection{Reflection Nebulae}
Our hierarchical models, with their overall spherical distribution of
clumps, are primarily aimed at interpreting reflected light from
reflection nebulae illuminated by an exciting star. We will discuss
reflection of the diffuse Galactic radiation field from dusty globules
at the end of this section.

We now examine the limits on $a$ imposed by the amount of scattered
light observed in reflection nebulae. Since real nebulae show the
illuminated cloud as it is viewed from only one direction, we must
consider the spread of (\tex, \tsca) from any particular dust
distribution. We seek \amax{} and \amin, the minimum and maximum
values of $a$ that can fit the observations.

For illustrative purposes, we consider the 1300 \AA{} observations of
\ngc\ with 50\% of the stellar extinction assumed to be within the
nebula (Witt et al. 1993; the circled dot in Figures \ref{fig1} and
\ref{fig2}). In Figure \ref{fig1} it lies in the middle of the
envelope of the (\tex, \tsca) points from various viewing angles. If
we decreased the $a$ of the model, the scattered flux at each
direction would become fainter and \tsca{} larger. The stellar fluxes
would be unchanged because they depend only on extinction, not
scattering. Thus, the pattern of points would move up in the diagram,
and the point could no longer be fitted. To find the \amin{} for
\ngc{}, we must vary \tz{} and initial seeds (dust distributions) and
find the minimum $a$ for which the envelope of individual points lies
just above the observation. A similar procedure leads to \amax. Of
course, it is improbable that the actual $a$ is as low as \amin{}
because the observed (\tex, \tsca) would probably not lie at the edge
of the (\tex, \tsca) distribution as seen from a random direction.

We have performed this exercise and found the following:

(a) All of our models required an albedo of $\gtrsim0.5$ to fit the
lower point of \ngc. Varying parameters and geometries to obtain the
absolute minimum $a$ (0.46; see below) produced only changes in
details. This result is robust because the observed \tex{} is near the
most efficient value for producing scattered light, as suggested by
the uniform models (see the minima of \tsca{} in the curves in Figure
\ref{fig1}).

(b) The scattered light is near its maximum when \tz{} $\sim$2, where
the mean \tex{} is about unity, so \amin{} is best estimated for our
hierarchical models near this value. The maximum scattered light for
uniform models occurs at \tz{} $\sim$ 1.1 -- 1.4, increasing slowly
with the albedo.

(c) The smallest \amin{} we found for the circled dot in Figure
\ref{fig1} is 0.48. It occurs for \tz{} $\sim$2, $D$ = 2.6, and a
particular seed whose (\mtex, \mtsca) point is low in Figure
\ref{fig4}. The effects of $g$ are minimal on the locus of the lower
envelope of the individual points; the minimum $a$ is achieved for
$g\ge0.5$. We tried such a wide range of models that we doubt that
there is an \amin{} drastically lower.

(d) No useful estimate of \amax{} is possible because inefficient
scattering can be produced by extreme clumping (as small a $D$ as
allowed) and with no uniform component. For $D$ = 2.3, an $a\sim1$ is
needed to fit \ngc{} for some distributions of clumps. For $D$ = 2.6,
\amax{} is about 0.8. The dust geometry that produced \amin{} did not
allow $a$ to be above 0.6, but it was chosen to have especially
efficient scattering. Alternatively, very optically thick models that
have low stellar extinctions as seen from some viewing angles can fit
observed fluxes with very high albedos. For instance, we have fitted
the \ngc{} point in Figure \ref{fig1} with \tpm, \tz{} = 8, $a$ =
0.85. We have no doubt that many optically thick hierarchical models
could fit the observations as well. Perhaps such extreme models could
be ruled out with far-infrared (FIR) fluxes, although the low
absorption tends to compensate for the high optical
depth. Furthermore, the FIR arises from dust surrounding the star in
all directions, while the scattering is mainly from dust close to
the line of sight. In any case, these models are hardly reasonable on
physical grounds.

{\it Voyager 2} and HUT observed \ngc{} at 1000\AA\ (Witt et
al. 1993).  Uniform models predict $a\sim$0.45 if 40\% of the stellar
extinction occurs within the nebula. Hierarchical models can fit the
observation (not plotted, but in the upper right corner of Figures 1
-- 3) with $a$ = 0.8 with $\tau_0\gtrsim4$. Of course, we make no
claim that such a high $a$ is correct. On the other hand, we can fit
the point with $a$ = 0.4, lower than the uniform models. Once again we
see that the observations are subject to a very ambiguous
interpretation.

Burgh, McCandliss, \& Feldman (2002) used a uniform model for
NGC\,2023 with \tex{} = 2.4. They derived an albedo of
$0.39^{+0.12}_{-0.05}$ at 1345\,\AA. Our hierarchical model shown in
Figure \ref{fig1}, with $a$ = 0.6, could accommodate their
observations, and others that scatter light inefficiently could have
$a\sim1$.

\begin{figure}
\plotone{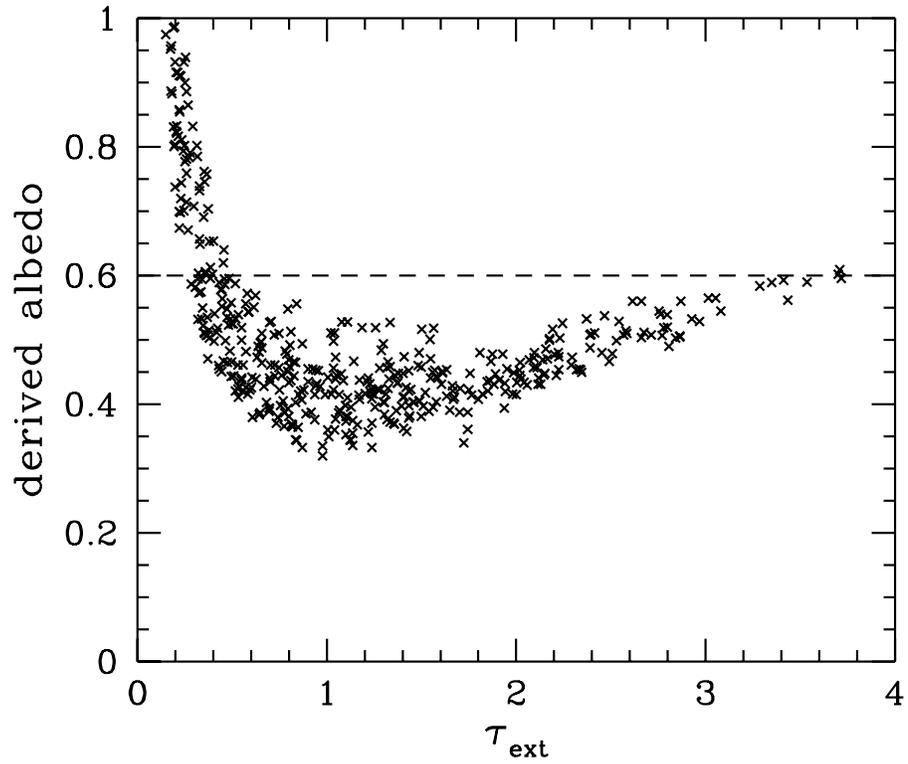}
\caption{
The albedo derived from uniform models, using the scattered flux and
extinction of a typical purely hierarchical model as seen from various
viewing angles, plotted against the stellar extinction. Parameters are
given in the text. The ``true'' albedo used in the generating model,
0.6, is marked by the dashed line.
\label{fig6} }
\end{figure}

Figure \ref{fig6} shows the albedos derived from uniform models, using
the scattering and extinctions from the hierarchical model
$(a,\,g,\,D,\,\tau_0)$ = (0.6,\,0.6,\,2.3,\,1), plotted against the
stellar extinctions at each viewing angle. The albedo used to generate
the scattered fluxes and extinctions, 0.6, is marked by the dashed
line. The distribution of points for the hierarchical model with \tz{}
= 2 is virtually identical. We see that the minimum $a$ derived from
uniform models is $\sim$0.32; the average for $0.6\le\tau_{\rm
ext}\le1.6$ is $\sim$0.42. At moderate extinction, leading to bright
scattered flux, uniform models underestimate $a$. At small \tex{}
uniform models overestimate $a$. A hierarchical model in which the
star happens to be within a clump has smaller errors. Adding a
constant density decreases the errors at \tex\ $\sim$ 1 so that the
average $a$ is $\sim $ 0.5. Of course, there is no error in the
unrealistic case of the density being uniform throughout.

\begin{figure}
\plotone{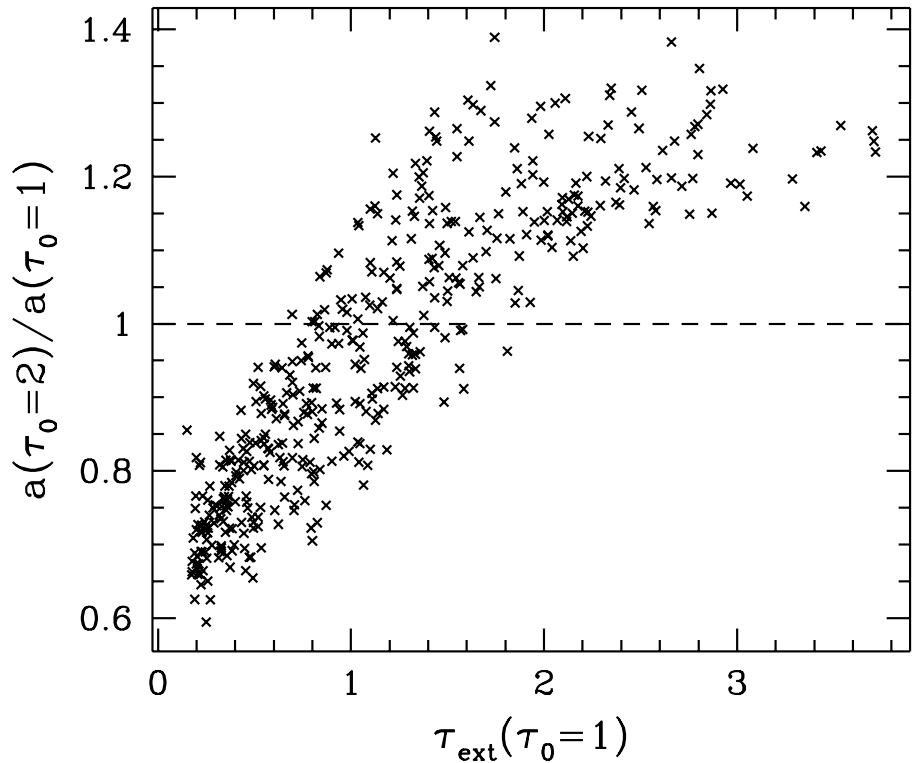}
\caption{ A test of whether {\it relative}
values of the albedo can be derived from observations of a nebula at
two different wavelengths or optical depths, with a given arrangement
of dust relative to the central star. Here we compare averaged optical
depths, \tz, of 2 and 1, for a particular hierarchical model. Since
only \tz{} is varied, the geometry of the dust is the same. The
scattering and stellar fluxes at various viewing angles were
interpreted with uniform density models. The ratio of the derived
albedos is plotted against the smaller stellar extinction. The dashed
line gives the ``true'' value of unity, since all points were derived
with $a$ = 0.6.
\label{fig7}
}
\end{figure}

An important question is, how reliably can we determine $a(\lambda)$
from observations of a particular nebula covering a significant range
in wavelengths, implying differing opacities or optical depths? In
this case, we might hope to find {\it relative} optical properties of
the grains accurately even with uniform models. Figure \ref{fig7}
shows the results for two wavelengths for which the extinctions (both
\tz{} and \tex) differ by a factor of 2: for example,
(1000\,\AA)/(1540\,\AA) if the extinction law is the Galactic average
(Cardelli, Clayton, \& Mathis 1989) and
$R_V=A(V)/[A(B)-A(V)]=3.1$. The figure displays the ratio of
$a$(\tz=2) to $a$(\tz{}=1), both albedos derived from uniform models,
for the same hierarchical model as shown in Figure \ref{fig1}. The
abscissa in the figure is \tex{} for the lower extinction. All points
in the figure were computed with the $a$ = 0.6, so the ``true'' value
of the albedo ratio is unity. We see that uniform models have
difficulty in predicting even relative values of $a(\lambda)$, since
the spread in $a(\tau_0=2)/a(\tau_0=1)$ is $\sim$40\%. The figure
gives the impression that if \tex{} is $\ge1.5$, uniform models would
derive a greater albedo at short wavelengths (say, 1000\,\AA) than at
longer (1540\,\AA\ for a factor of 2 in the extinction). For \ngc{}
the opposite is found to be the case (Murthy et al. 1993; Witt et
al. 1993). From Figure \ref{fig7} we would conclude that
$a$(1000\,\AA)$<a$(1540\,\AA) is robust, since Witt et al. (1993)
suggest \tex(1540\,\AA) $\sim$(1.5 -- 2). However, models with the
star embedded in a clump have the albedo ratios shift from mostly $<1$
to mostly $>1$ at \tex(1540\,\AA) $\sim2.5$, considerably greater than
shown in the figure. These models would produce an illusory result
$a$(1000\,\AA) $<a$(1740\,\AA).

The basic problem is that the importance of a clump depends upon its
optical depth. If a clump becomes optically thin, its existence no
longer very significant to the radiative transport. Similarly,
increasing the optical depth of an optically thick clump has a limited
effect on the radiation. While the density structure within the ISM is
independent of the wavelength, its geometry as regards radiation
transfer is not. Since the clumping produces the large spread in
\tsca($\theta,\,\phi)$, it is not surprising that there is a wide
spread in the importance of the clumps among various viewing angles.

The problems of determining the variation of albedo at various
wavelengths are small if the ratio of optical depths is close to
unity. Statements that the albedo is smaller at the 2175 \AA\ feature
than on either side of it (e.g., Witt et al. 1992) seem robust,
besides being completely plausible physically. At nearby wavelengths
$A(\lambda)/ N($H), the extinction per H nucleus, can be the same at
three wavelengths, and comparison of the relative albedos between
those wavelengths is completely safe.

We have seen that several properties of reflection nebulae change if
the exciting star is embedded within a rather dense clump of
dust. There is an increase in the scattered light; the lowest values
of \mtsca{} for purely hierarchical models in Figure \ref{fig4}$a$
(i.e., the large amounts of scattered light for the purely
hierarchical models) belong to such cases. The central clump ensures
that the angle-averaged extinction, \mtex, is larger than for most
models with the same \tz\ (i.e., the points in Figure \ref{fig4}$a$
tend to have large \mtex{} if \mtsca{} is small). Figure
\ref{fig4}$b$ shows that the addition of a constant density component
makes the effects of a central clump rather unimportant because the
radiation scattered by the uniform dust dominates. Another effect of a
star being embedded in a dense clump is that the errors of the albedo
derived from observations of the same object at different wavelengths,
illustrated in Figure \ref{fig6}, are reduced because more of the
scattering arises from the central clump. For such a clump, the
opacity, but not the geometry, changes with wavelength. Finally, there
would tend to be more FIR radiation from an embedded star because the
grains are relatively close to the star, so they are absorbing a
relatively large stellar flux, and the mean temperature of the grains
would be relatively large for the same reason.

Galactic radiation scattered by isolated globules seen away from the
Galactic plane provides another diagnostic of grains (Mattila
1980). Witt, Oliveri, \& Schild (1990) have analyzed such a globule
and give references to other examples and discussion of the
process. The density is concentrated towards the center of such
globules. The reflected intensity rises outward from the center,
peaks, and falls off to the sky brightness at the edge. showing that
the scattering is forward-throwing. The reflected intensity can be
related to the incident radiation predicted for both stars and the DGL
out of the Galactic plane. Results were that at 4700 \AA\ grains are
forward-throwing ($g\sim$0.8) and very efficient at scattering
($a\sim0.8$). The globule was modeled with a very centrally condensed
density distribution that is smooth and spherically symmetric, as
seemed completely appropriate at the time. Since turbulence driving
clumps in the diffuse ISM may decay when driving into a steepening
density gradient, perhaps the ISM within globules is relatively free
from clumping. We cannot assess the uncertainties in the grain
properties derived from globules until we know more about the clumping
within them. The major uncertainty would then be the prediction of the
radiation incident upon them from the hierarchically clumped Galaxy.

\subsection{The Diffuse Galactic Light}
There is no doubt about the applicability of hierarchical clumping to
the DGL. Observations such as those at 1740 \AA\ (Schiminovich et
al. 2001) are based on large portions of the sky. We have seen that
uniform spherical models of reflection nebulae underestimate $a$ if
applied to optically thick situations (large \tz, but not necessarily
large \tex). Our reflection nebula models are not ideally suited for
interpreting the DGL in the UV, since distributions of the dust and
exciting stars are skewed towards the Galactic plane. However, our
results have strong implications as regards the interpretation of the
DGL as discussed in Henry (2002, hereafter H02). In that paper it was
concluded that $a\lesssim0.1$ for $\lambda\lesssim1400$ \AA.

H02 performed radiative transport calculations by using $B, V$
magnitudes and the spectral type for each star in the entire
Hipparchos Input Catalog. The reddening, $E(B-V)$, follows from
$(B-V)$ and spectral type. The extinction $A_\lambda$ is
$E(B-V)\,R_V\,(A_\lambda/A_V)$, where $R_V$ is usually taken to be
3.1. The \tex{} for each star is $A(\lambda)$/1.086.

The heart of H02 is its unique treatment of radiative transfer. The
scattering is taken to be
\begin{equation}
\exp(-\tau_{\rm sca})=a(1-e^{-\tau_{\rm ext}})
\exp\{-(1-a)[\tau_{\rm ext}+\ln(1+e^{-\tau_{\rm ext}})-\ln2]\}\ ,
\end{equation}
\noindent independently of the geometry. However, Witt, Friedmann, \&
Sasseen (1997) have shown from clumpy models that the distribution of
cloud optical depths can strongly affect the interstellar radiation
field.

At low optical depths, the H02 relation between \tex{} and \tsca{}
approaches the uniform model. The dotted lines in Figure \ref{fig1}
show the H02 models for three albedos identified by the nearby solid
lines. We see that the H02 assumption predicts more scattered light
than uniform models, and a lower albedo follows.

For \lam $\lesssim2000$ \AA, the \tex\ for most hot stars is probably
rather large, since the mean $A(V)$ is $\sim$1 magnitude kpc$^{-1}$
and the ultraviolet is 3 -- 5 times larger. For \tex{} = 1.5, Figure
\ref{fig1} shows that the H02 model with $a$ = 0.6 predicts a
scattered flux $\sim$exp(0.4) = 1.5 times larger than that ``typical''
hierarchical model, chosen for illustrative purposes only. The
conclusion of H02 that $a$(1000\,\AA) is very low (``perhaps 0.1'') is
perhaps understandable. Such a low albedo is not compatible with the
observations of \ngc{} by Witt et al. (1993), in which the scattered
flux was larger than the stellar, or Murthy et al. (1993), which
suggests that $a$(1150 \AA) $\sim 0.55$ if 40\% of the extinction is
within the nebula, as assumed by Witt et al. (1992).

Murthy, Henry, \& Holberg (1991) used {\it Voyager\,2} to search for
DGL for $\lambda\le1300$\,\AA. They failed to detect any, possibly
confirming the H02 picture. Our reflection nebula results commonly
show that the same nebula can show very low and rather high scattered
fluxes as seen from different angles. We strongly suspect that the
same property applies to the DGL. Our general result from hierarchical
models is that photons, especially the unscattered ones coming
directly from the star, can leak out of dust distributions far more
easily than either uniform-density or H02 models would predict, so a
dark sky does not necessarily imply a low albedo.

\section{Final Remarks and Summary}
We have arrived at a pessimistic, but we feel realistic, assessment of
the ability of reflection nebulae illuminated by embedded stars to
serve as diagnostics of interstellar grain properties. It is better to
realize what we do not know, rather than believe that we know more
than we do! Even if the scattered light from a reflection nebula and
the extinction suffered by its exciting star are observed perfectly,
the geometry of the dust is vital in the interpretation of the
observations. Hierarchical clumping, ``clumps within clumps within
clumps'', has a very large impact on the interpretation of both
reflection nebulae and the DGL as compared to smooth density
structures. Hierarchical structures, demonstrated by CO, 100 $\mu$m,
various other molecules, and H~I observations, make any determination
of the optical properties of grains from reflection nebulae very
difficult. Our models had plausible power spectra for the
autocorrelation function of the projected densities, with an exponent
of the power of $\sim-3$, unlike either uniform models or those that
are clumped on a single scale (exponent $\sim$0). The exact geometry
of the dust is very difficult to determine because it depends on the
placement of hierarchical clumps of dust around the star.

We have made no interpretations of any reflection nebulae.  It is now
relatively easy to use Monte Carlo transfer codes to predict the
scattering (and polarization) for objects {\it with a known geometry.}
Inverting observed intensities to determine the correct geometry,
though, is {\it very} difficult, even with observations with good
spatial resolution taken in multiple wavelengths. Our hierarchical
models have been schematic and do not begin to exhaust the
possibilities of the real geometry of any reflection nebula.
Observations in the infrared and line emissions with good spatial
resolution may eventually put useful limits on the actual geometry of
the gas and dust, but the problem of inverting the data will always be
difficult.

An interesting possibility is that isolated Bok globules may not be
strongly clumped, so reflection of Galactic light from them can
diagnose grain properties. The penetration of turbulence into them
might be weaker than its percolation through larger
structures. Nonclumpy globules can serve to diagnose grains because
they scatter incident Galactic radiation (see Witt et al. 1990 for
references and an explanation).

Each of our hierarchical models shows a very wide range of extinction
optical depths (\tex) and scattering (\tsca) when the nebula is viewed
from various directions. The \tex{} and \tsca{} observed for a
reflection nebula represent a view of the dust from one direction.
There is a very wide range of optical properties ($a,\,g$) that can
fit a given observation. With one or more hierarchical models we can
fit the best observed UV reflection nebula, \ngc, with albedos $\ge$
0.5 over the range of $g$ that we tested (0.6 -- 0.85) and with a
variety of averaged optical depths through the model.

In general, reflection nebulae have substantial optical depths (\tex\
$\sim1-3$); otherwise, the scattered fluxes are too faint. At
substantial optical depths, uniform models can underestimate the
albedo rather severely because they overestimate the scattered
light. Unless the star happens to be embedded within a dense clump,
radiation can leak out of hierarchical clumps much more easily than
from uniform dust.

The Diffuse Galactic Light (DGL) has been interpreted (Henry 2002)
with a recipe that predicts much more scattering from each star than
our models, resulting in a low estimates of the UV albedo. Murthy et
al. (1991) failed to detect UV DGL, but our fractal models often have
many directions from which the star and scattered light are very
faint. Thus, we do not believe that the faintness of the DGL in
particular directions necessarily signifies a low albedo.

The determination of both $a$ and $g$ depends upon the variation of
reflected intensity across the face of a centrally illuminated
reflection nebula, while we have only considered the flux of the
scattered light. We feel that if the flux is as weakly constrained as
our models show, the intensity will be similarly subject to variation
because of the unknown placement of the star relative to the
surrounding dust.

Even the relative variation of albedo with wavelength is difficult to
determine from reflection nebulae if there are significant changes in
opacity among the wavelengths. The problem is that the geometry is not
really the same when the opacity changes. As opacity per H atom
increases, optically thin clumps become thick and scatter light with a
different geometrical arrangement, so the geometry of the nebula
depends on wavelength.

Our overall assessment is that the optical properties of grains are
probably as well constrained by theory as by observations. This
statement is made in spite of well-known uncertainties in the theory
of interstellar grains. These uncertainties are major: whether typical
large grains are chemically homogeneous (e.g., silicate or
carbonaceous) or composite, whether grains contain voids, or if grains
have very loose (``fractal'') structure. However, theory does not
permit the large range of possible albedos provided by reflection
nebulae. At first glance, various theories seem quite
different. Weingartner \& Draine (2001) have PAH molecules, small
grains, and large grains of silicate or graphite. Li \& Greenberg
(1997) have PAHs, small grains, and large grains with silicate cores
and organic refractory mantles. Mathis (1996) has composite grains
with silicates, carbon, and vacuum intermixed within the same grain,
plus small graphite that mimics PAHs. Draine \& Lee (1984) have
silicate and graphite with a truncated power-law distribution of
sizes. Each size distribution is different. They all predict roughly
the same $a(\lambda)$ because the response of the grains to radiation
is set by fitting the well-observed interstellar extinction law
\tex($\lambda$), and cosmic abundances (though somewhat controversial
themselves) constrain the materials. For instance, the values
predicted for $a$(1430 \AA) are (0.40, 0.30, 0.37, 0.40) for
Weingartner \& Draine (2001), Li \& Greenberg (1997), Mathis (1996),
and Draine \& Lee (1984), respectively. The $A(V)$ are (0.65, 0.63,
0.53, 0.57).

Fortunately, optical depths and column densities derived from
comparisons of absorption and scattering along the same very narrow
beam (i.e., towards a star) are not affected by the microstructure
within the ISM. These measurements include most determinations of
depletions of various ions in the ISM, as well as the wavelength
dependence of the interstellar extinction law. Other implications for
the hierarchical nature of dust clumping include a greatly increased
penetration of radiation into molecular clouds and PhotoDissociation
Regions (PDRs).

\acknowledgements We would like to thank Prof. A. Witt for constructive
comments on a draft of this paper that improved it considerably. BW
and KW acknowledge financial support from NASA's Long Term Space
Astrophysics Research Program, NAG5~6039 (KW), NAG5~8412 (BW), the
National Science Foundation, AST~9909966 (BW and KW), and a PPARC
Advanced Fellowship (KW). We appreciate the constructive and prompt
comments of two anonymous referees. This research has benefited from
the NASA Astrophysics Data System and the SIMBAD database.

\newpage

\end{document}